\documentclass[%
 reprint,
superscriptaddress,
 amsmath,amssymb,
 aps,
pra,
]{revtex4-2}

\usepackage[normalem]{ulem}

\usepackage{graphicx}
\usepackage{dcolumn}
\usepackage{bm}
\usepackage{etoolbox}
\usepackage{hyperref}
\usepackage{lipsum}

\usepackage{xcolor}
\usepackage{threeparttable}

\usepackage{multirow}

\usepackage{float}

\begin{document}

\setlength{\tabcolsep}{6pt}
\renewcommand{\arraystretch}{1.1}

\title{Ab Initio Calculations of the Static and Dynamic Polarizability of BaOH}

\author{Eifion H. Prinsen}
\affiliation{Van Swinderen Institute for Particle Physics and Gravity, University of Groningen, The Netherlands}
\affiliation{Nikhef, National Institute for Subatomic Physics, Amsterdam, The Netherlands}

\author{Anastasia Borschevsky}
\author{Steven Hoekstra}
\affiliation{Van Swinderen Institute for Particle Physics and Gravity, University of Groningen, The Netherlands}
\affiliation{Nikhef, National Institute for Subatomic Physics, Amsterdam, The Netherlands}

\author{Achintya K. Dutta}
\affiliation{Department of Chemistry, Indian Institute of Technology Bombay, Mumbai 400076, India}

\author{Sudipta Chakraborty}
\affiliation{Department of Chemistry, Indian Institute of Technology Bombay, Mumbai 400076, India}

\author{Bart J. Schellenberg}
\affiliation{Van Swinderen Institute for Particle Physics and Gravity, University of Groningen, The Netherlands}
\affiliation{Nikhef, National Institute for Subatomic Physics, Amsterdam, The Netherlands}

\author{Luk\'{a}\v{s} F. Pa\v{s}teka}
\email{l.f.pasteka@rug.nl}
\affiliation{Van Swinderen Institute for Particle Physics and Gravity, University of Groningen, The Netherlands}
\affiliation{Nikhef, National Institute for Subatomic Physics, Amsterdam, The Netherlands}
\affiliation{Department of Physical and Theoretical Chemistry, Comenius University, Bratislava, Slovakia}

\author{I. Agust\'{\i}n Aucar}
\email{agustin.aucar@conicet.gov.ar}
\affiliation{Van Swinderen Institute for Particle Physics and Gravity, University of Groningen, The Netherlands}
\affiliation{Nikhef, National Institute for Subatomic Physics, Amsterdam, The Netherlands}
\affiliation{Instituto de Modelado e Innovaci\'on Tecnol\'ogica (UNNE-CONICET), Facultad de Ciencias Exactas y Naturales y Agrimensura, Universidad Nacional del Nordeste, Corrientes, Argentina}

\date{\today}

\begin{abstract}
We present \textit{ab initio} calculations of the static and dynamic electric dipole polarizability of the barium monohydroxide ($^{138}$BaOH) molecule, using relativistic coupled-cluster theory. By thoroughly investigating the dependence of the calculated polarizabilities on computational parameters (basis set size, treatment of relativistic and QED effects, level of description of electron correlation, and vibrational corrections), a procedure for determining uncertainties is constructed and applied. The electric dipole moment of BaOH is also calculated and compared to experiment. The static and dynamic (for $\lambda=1064\;\text{nm}$) polarizability tensor elements were calculated within the electronic ground state for both the (000) vibrational ground state and the (010) bending vibrationally excited state, the latter state being particularly interesting for a wide range of quantum experiments. The ground-state static polarizability tensor elements were calculated to be 200.8(24)\,a.u.\,and 297(5)\,a.u.\,for the parallel and perpendicular components, respectively.
\end{abstract}

\maketitle

\section{Introduction}\label{sec:introduction}

Accurate knowledge of molecular properties is crucial for designing experiments in both physics and chemistry \cite{carr_2009}. Where experimental data for a particular property is unavailable, theoretical predictions can provide guidance instead. In order to be reliable and useful in an experimental context, such predictions should be based on accurate calculations and be accompanied by meaningful uncertainties.

In recent years, di- and triatomic molecules have become increasingly popular for high-precision experiments testing the validity of the Standard Model (SM) \cite{aggarwal_2018,acme_2018,hutzler_2020,Roussy2023,anderegg_polyatomics_2023,bause2024,arrowsmithkron_2024,ciamei_2025}, as molecular electronic structure may provide strong enhancement of charge ($\mathcal{C}$), parity ($\mathcal{P}$), or time-reversal ($\mathcal{T}$) symmetry-violating effects \cite{Safronova2018}. One of such $\mathcal{CP}$-odd effects arises from the interaction between the electron electric dipole moment (eEDM) of the unpaired electron and the internal electromagnetic fields in a molecule.
The eEDM serves as an important benchmark of the SM and beyond-the-Standard Model (BSM) theories \cite{Chupp2019}. While the SM prediction of the eEDM, of order $10^{-40}\;e\,\text{cm}$ \cite{Yamaguchi2020}, is still far from experimental reach, BSM theories generally predict larger values. Significant progress was made in recent years constraining the eEDM, with the current upper bound set at $2.1\times10^{-29}\;e\,\text{cm}$ in a combination of experiments on HfF$^+$ and ThO molecules \cite{Roussy2023,acme_2018}.

The NL-eEDM collaboration has recently proposed a next-generation experimental setup making use of laser-cooled \cite{vanhofslot_2025} and trapped BaOH molecules \cite{bause2024}. BaOH was identified as a promising candidate due to its favorable properties, including the possibility to laser-cool and trap this molecule, as well as its strong sensitivity to the eEDM~\cite{Denis2019,Gaul_2020,Cheng_2021} owed to its open-shell electronic structure. Moreover, by polarizing BaOH using a tunable electric field in the (010) vibrational bending mode, it is possible to obtain a near-zero $g$-factor spin state \cite{anderegg_polyatomics_2023,bause2024}. Such a state significantly suppresses the systematic sensitivity to magnetic fields, whilst preserving the sensitivity of the molecule to the eEDM. The two bending modes of BaOH are degenerate due to its linear geometry, endowing its rotational spectrum in a bent state with so-called $\ell$-doublets. These allow polarization of the molecule at relatively small electric fields \cite{hutzler_2020}.

Several polyatomic species, such as CaOH and SrOH, have already been successfully laser-cooled \cite{mitra_2020,AugAndHal23}, loaded into a magneto-optical trap \cite{baum_2020,vilas_2022}, and even transferred to an optical dipole trap (ODT) \cite{Hallas2023,SawNasLun25}. This process, however, remains challenging due to the complex level structures that are involved. Consequently, highly accurate knowledge of several molecular properties is required to further develop existing cooling and trapping techniques \cite{chen_2016,hao_2019,caldwell_2020}.

The electric dipole polarizability characterizes the response of a molecule to an external uniform electric field, either static (static polarizability) or time-dependent (dynamic polarizability), resulting in a shift of its energy levels. Using the polarizability, it is possible to estimate the potential and trap depth corresponding to an ODT. When the trapping frequency is very far detuned from transitions in the molecule, the static polarizability provides a reasonable estimate of the dynamic response \cite{Grimm1999,singh2023}. However, especially in the vicinity of a transition, the frequency-dependent dynamic polarizability is required. For trapped BaOH, the use of near infrared ($\lambda=1064\,\text{nm}$) light to form the ODT gives a good balance between sufficient detuning and practical considerations, such as the availability of sufficiently powerful and stable lasers. A well-chosen laser detuning also determines the balance between trap depth and residual photon scattering rate.

To accurately predict these static and dynamic polarizabilities, highly reliable electronic-structure methods are essential. Over the past decades, the coupled-cluster (CC) approach has emerged as one of the most powerful and trusted \textit{ab initio} methods for predicting atomic and molecular properties. Consequently, in the present work, this approach has been employed to provide reliable theoretical predictions of the static and dynamic electric dipole polarizability of BaOH.

Molecular response properties---such as electric dipole polarizabilities, which are the focus of the present paper---can be investigated within the CC framework either via finite-field (FF) techniques or through linear-response (LR) methodologies based on analytical energy gradients, developed originally within the nonrelativistic (NR) Schrödinger formalism in the 1970s~\cite{Monkhorst77,Dalgaard83,Jorgensen88,Helgaker89,Koch1990}. In this latter context, the first NR-LR-CC calculations of frequency-dependent dipole polarizabilities were reported by Kobayashi \textit{et al.}~\cite{Kobayashi94}. Since then, this methodology has been widely employed for the study of response properties, at the CC single and double excitation level (CCSD). Later, this was extended to include triple and higher excitations (see, for example, Refs.~\citenum{Gauss1997}, \citenum{Gauss2000}, and \citenum{Gauss2000bookchap} and citations therein).

However, the NR treatment becomes insufficient when dealing with heavy elements such as barium, for which relativistic effects contribute significantly to the electronic structure and, consequently, to response properties. The accurate description of heavy atoms and heavy-element-containing molecules, therefore, requires incorporation of relativistic effects.
From a semi-relativistic perspective, the combined influence of electron correlation and relativistic effects on static electric properties, including dipole polarizabilities, has been systematically explored by Sadlej, Kellö, Urban, and co-workers across a wide range of molecular systems using \textit{ab initio} FF methodologies~\cite{Kello1990,Kello1995,KelloJCP1995,Kello1996,Ilias03,Urban2005,Holka2009}. Their studies were conducted within quasi-relativistic and scalar-relativistic frameworks (mostly based on the Douglas--Kroll--Hess approach), supplemented with \textit{a posteriori} spin-orbit corrections. Moreover, they developed specialized POL basis sets tailored for accurate calculations of electric properties~\cite{Sadlej88,Sadlej92,Sadlej92b,Sadlej96,Miadokova97,Cernusak03}.

It is worth noting that the static and dynamic electric dipole polarizabilities of atomic systems have been extensively investigated at NR, semi-relativistic, and relativistic levels of theory (see, e.g., Refs.~\citenum{Lim1999}, \citenum{Sahoo2007}, and \citenum{Mitroy2010}). 
In contrast, (four-component) relativistic CC molecular polarizability calculations are rather scarce.
For example, Thakur et al.~\cite{Thakur2024} recently employed the FF technique to study this property in a series of ionized molecular systems. 
Very recently, new implementations for calculating relativistic response properties using analytical gradient techniques at the CC level of theory have been developed independently by Yuan \textit{et al.}~\cite{Yuan_RCC-LR_2024,Yuan_RCC-QR_2025} and Chakraborty \textit{et al.}~\cite{Chakraborty-jpca-2025,chakraborty2025lowcost}.

In this work, we present \textit{ab initio} calculations of the static and dynamic dipole polarizabilities of BaOH that are systematically converged. The four-component relativistic coupled-cluster approach, complemented by very large basis sets, was used in the calculations of the static polarizabilities by applying the FF method. Despite the recent developments of relativistic LR-CC implementations~\cite{Yuan_RCC-LR_2024,Chakraborty-jpca-2025,chakraborty2025lowcost}, these do not include orbital relaxation effects. For this reason, striving to obtain the most accurate results possible, we have applied numerical gradients instead. Nevertheless, we present the differences between the results obtained with the (orbital-relaxed) FF method and others based on analytical gradients with unrelaxed orbitals. Furthermore, we have determined the theoretical uncertainties of the calculated static polarizabilities based on an extensive computational study.

The dynamic polarizability was calculated at various frequencies of the external electric field, including the frequency corresponding to the transition used for the laser-cooling and trapping. These calculations were also performed within the coupled-cluster approach, with relativistic effects incorporated through an effective core potential (ECP).
By indirectly incorporating the more accurately determined static polarizability, the uncertainty of the dynamic polarizability at the desired frequency is reduced.

Previous experimental work on BaOH includes spectroscopy of several electronic, vibrational, and rotational transitions \cite{KinseyNielsen_1986, Anderson1993, Fernando1990, Tandy2009, Pooley1998}. A number of recent theoretical studies of this molecule are also available, focusing predominantly on its potential for precision measurements \cite{Denis2019, Gaul_2020, Cheng_2021}.

\section{Theory}

The polarizability tensor describes the distortion of the molecular electronic distribution in an external electric field. In this work, the static polarizability $\alpha(0)$ of BaOH is determined using the FF method \cite{Cohen1965}, while the dynamic polarizability, $\alpha(\omega)$, is obtained by solving LR equations within the non-relativistic (NR) polarization propagator theory \cite{Oddershede1984,Olsen1985,Oddershede1987}.

In the FF method, the electronic energy $E(F_q)$ of a system relative to the energy $E(0)$ in the absence of the field is expanded in terms of the external electric field strength $F_q$, oriented along axis $q$. In the molecular frame, we have:
\begin{equation}\label{eq:finite_field}
\begin{split}
    E(F_q) &= E(0) + \frac{d E}{d F_q}\bigg|_{0}F_q + \frac{1}{2}\frac{d^2E}{d F_q^2}\bigg|_{0}F_q^2 + \ldots\\
    &\equiv E(0) -\mu_{q} F_q - \frac{1}{2}\alpha_{qq} F_q^2 - \ldots\;\;,
\end{split}
\end{equation}
where $\mu_{q}$ is the electric dipole moment in that frame, and $\alpha_{qq}$ is the electric dipole polarizability of the system, both along the $q$-axis, defined as the first and the second field-derivative of the energy, respectively:
\begin{equation}\label{eq:alpha-FF}
     \mu_{q} = -\frac{dE}{dF_q}\bigg|_{F_q=0}
     , \qquad \alpha_{qq} = -\frac{d^2E}{dF_q^2}\bigg|_{F_q=0} .
\end{equation}
For a linear molecule oriented along the $z$-axis, we distinguish between the parallel and perpendicular components of the polarizability tensor, $\alpha_\parallel$ and $\alpha_\perp$ respectively:
\begin{equation}
\alpha_{zz} \equiv \alpha_\parallel,\;\; \alpha_{xx} = \alpha_{yy} \equiv \alpha_\perp\;.
\end{equation}

\section{Computational details}\label{comp}

The body-fixed electric dipole moment and static dipole polarizability were calculated using the relativistic single-reference coupled-cluster (CC) approach with single, double, and perturbative triple excitations (CCSD(T)) within the FF approach, using the DIRAC23 program package \cite{dirac-paper,DIRAC23}. In this finite-difference approach, the applied field strengths were chosen to be sufficiently large to avoid numerical problems and small enough to remain within the linear response regime, ensuring the numerical stability of the obtained results. The perturbation was added to the unperturbed Hamiltonian before performing the iterative SCF process, so that the results obtained include orbital relaxation effects. To reduce the computational costs, the exact 2-component (X2C) Hamiltonian \cite{Ilias2007} was used instead of the 4-component Dirac--Coulomb (DC) Hamiltonian.

The dynamic polarizability was calculated by solving the NR LR equations as implemented in the CFOUR program package \cite{CFOUR} at orbital-unrelaxed CCSD level \cite{Kallay2006}. To incorporate relativistic effects, an ECP \cite{Lim2006} on the heavy barium atom was used, replacing the inner core of 46 electrons. The calculations were performed for various frequencies of the external electric field, from the static limit ($\omega=0$) up to just beyond the first and second electronic transition frequencies of the molecule.

\begin{figure}[!hbpt]
    \centering
    \includegraphics[width=0.7\linewidth]{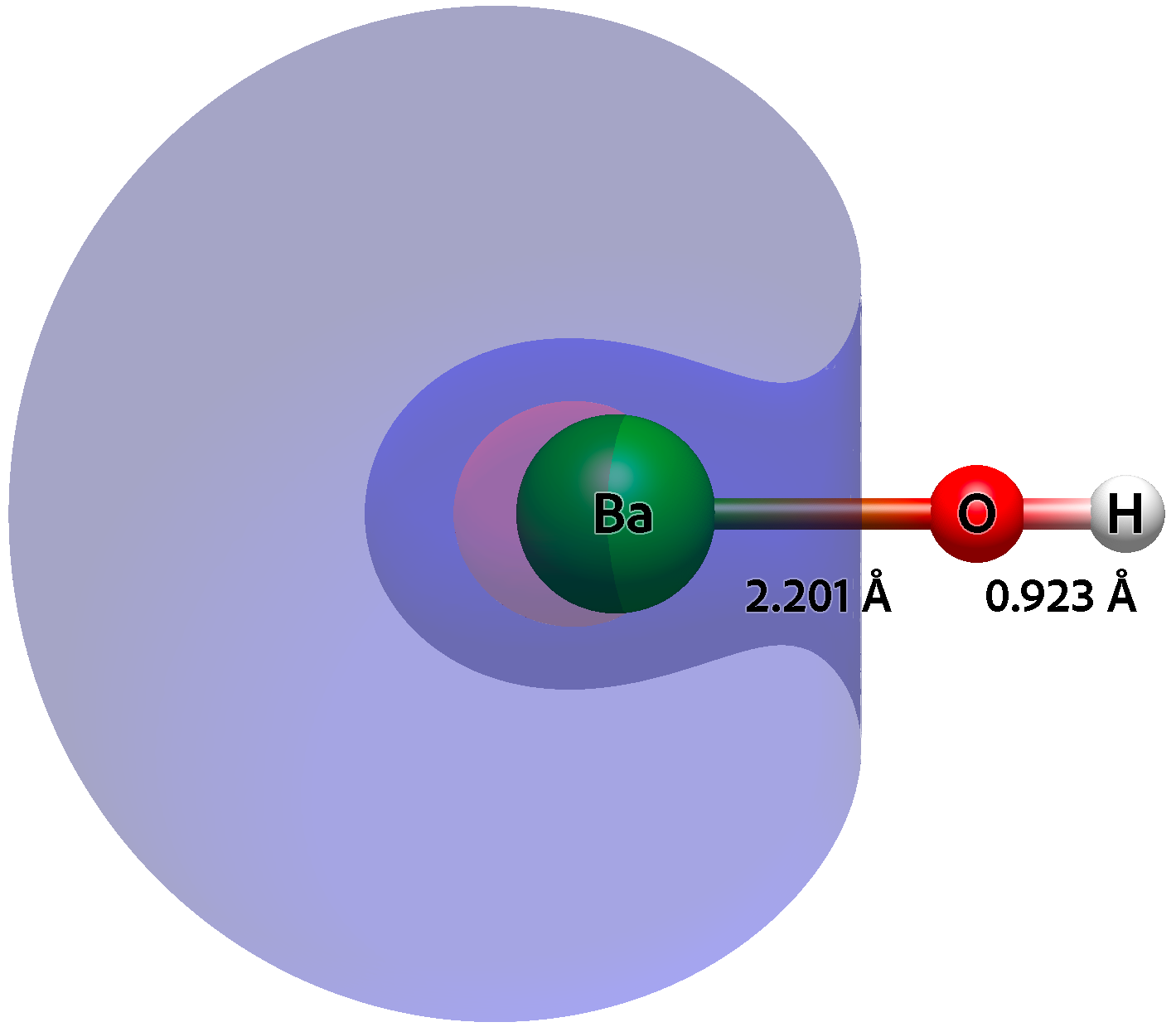}
    \caption{Linear geometry of BaOH and the isocontour representation of the singly occupied valence orbital ($\approx$ Ba:6s).}
    \label{fig:BaOH}
\end{figure}

All calculations were performed using the experimentally determined ground-state geometry of the linear BaOH molecule \cite{KinseyNielsen_1986}, with \mbox{$r(\text{Ba}$--$\text{O}) = 2.201 \;\text{\AA}$}, \mbox{$r(\text{O}$--$\text{H}) = 0.923 \;\text{\AA}$} (see Figure~\ref{fig:BaOH}).

\subsection{Static polarizability}

Singly, doubly, and triply augmented, as well as non-augmented valence (uncontracted) Dyall basis sets \cite{Dyall2009, Dyall2016} of double-, triple-, and quadruple-$\zeta$ quality were used in the calculations. The augmented basis sets are denoted as $n$-aug-dyall.v$X$z, where $n$ is s, d, or t, corresponding to the addition of one, two, or three layers of diffuse functions, respectively, added in an even-tempered manner, and $X$ is 2, 3, or 4, indicating the cardinality of the basis set.

For the static polarizabilities, we investigated the effect of the treatment of relativity by comparing results obtained using NR, X2C, and DC Hamiltonians. The Breit interactions were not included in the latter two Hamiltonians, and their effects will instead be estimated based on the literature. To estimate the effects of quantum electrodynamics (QED) on the polarizability, calculations were performed with the X2C Hamiltonian with self-consistently added Uehling \cite{Uehling1935} and Flambaum-Ginges \cite{Flambaum2005} effective potentials, corresponding to vacuum polarization and electron self-energy effects, respectively \cite{sunaga2022}.

In the CCSD and CCSD(T) calculations, 35 electrons were correlated by setting the active correlation space energy range from --20\,$E_\text{h}$ to 20\,$E_\text{h}$. Calculations with larger active spaces, --200\,$E_\text{h}$ to 200\,$E_\text{h}$ (59 electrons correlated) and --2000\,$E_\text{h}$ to 2000\,$E_\text{h}$ (all 65 electrons correlated), were carried out to estimate the uncertainty associated with the active space truncation in the CC procedure.

To calculate the electronic contribution to the body-fixed electric dipole moment and the static dipole polarizability within the FF approach, the perturbed energy of the molecule was calculated with the following electric field strengths (in a.u.): $0.0$ and $\pm0.0001$, oriented either parallel or perpendicular to the molecular axis, to determine $\alpha_{\parallel}(0)$ and $\mu_z$, and $\alpha_{\perp}(0)$, respectively. Note that the field perpendicular to the internuclear axis reduces the cylindrical $C_{\infty\text{v}}$ symmetry of the electronic wavefunction to $C_\text{s}$ symmetry.

The dipole moment vector in the molecular frame and the polarizability tensor elements were extracted by approximating the first- and second-order derivatives of the energy (see Eq.~\eqref{eq:alpha-FF}) at $F=0$ using the {two- and three-point} finite difference formulae, respectively.
Higher values of the field of $\pm 0.0003$ or $\pm 0.0005$ a.u. were used in cases where numerical noise dominates, which is more likely to occur for the second-order derivatives.

\subsection{Dynamic polarizability}\label{sec:comp-dyn}
We used Dunning's correlation-consistent cc-pV$X$Z-PP basis sets \cite{Hill2017, Dunning1992}, with $X$ = Q and 5, corresponding to quadruple- and quintuple-$\zeta$ quality, respectively, for calculating the dynamic polarizability tensor elements. The effect of basis set augmentation was investigated by performing calculations with non-, singly-, and doubly-augmented basis sets, the latter two denoted with an aug- and 2-aug- prefix, respectively.

The inner 46 core electrons of barium were replaced by an ECP~\cite{Lim2006} to model scalar relativistic effects. The remaining 19 electrons of BaOH were included in the CCSD active space.

The dynamic polarizability was calculated at several frequencies of the external electric field, ranging from zero frequency (corresponding to the static polarizability) to beyond the transitions to the $A^2\Pi$ and $B^2\Sigma$ states, i.e., from $\omega=0$ to $\omega=600$ THz. The static limit of the dynamic polarizabilities will be referred to as $\alpha_{\text{ECP}}(0)$ in the rest of this work. The dynamic polarizability values are shifted such that $\alpha_{\text{ECP}}(0)$ matches the corresponding $\alpha_{\parallel}(0)$ or $\alpha_{\perp}(0)$. A least-squares fit through the data was used to interpolate the polarizability at the desired field frequency of 282 THz (corresponding to a wavelength of 1064 nm, and a wavenumber of $\sim 9400$ cm$^{-1}$).

\subsection{Vibrational corrections}
Within the Born-Oppenheimer approximation (BOA), vibrational effects are not included when solving the electronic problem for clamped nuclei, and molecular properties are determined in this context only for the equilibrium geometry in the molecular frame. To find the polarizability of BaOH in the (010) bending vibrational state, we estimated the vibrational effects on the calculated polarizabilities, within the BOA, for the four vibrational modes. The corresponding geometry optimization and vibrational analysis were performed using the Gaussian program \cite{Gaussian}. 

The following equilibrium geometry was found at the CCSD level of theory using the aug-cc-pVQZ-PP basis set with corresponding ECP: $r(\text{Ba}$--$\text{O}) = 2.3925 \;\text{\AA}$, $r(\text{O}$--$\text{H}) = 0.9528 \;\text{\AA}$. 

The frequency analysis was subsequently carried out at the optimized geometry at the same level of theory, yielding two bending and two stretching modes. The bending modes are degenerate by symmetry and are referred to as $\Pi$. The two stretching modes are labeled $\Sigma_1$ and $\Sigma_2$, corresponding to the symmetric and anti-symmetric stretch, respectively. The $\Pi$ mode reduces the $C_{\infty\text{v}}$ symmetry of the electronic structure in the molecular frame to $C_\text{s}$ symmetry, resulting in $\alpha_{xx} $ and $\alpha_{yy}$ no longer being equal, and the polarizability tensor gaining non-zero off-diagonal components. The off-diagonal elements remain negligibly small and are thus omitted.

The polarizabilities (static and dynamic, the latter at the laser wavelength of $1064$ nm) and the single-point energies of BaOH were calculated at 10 displaced nuclear configurations along the normal mode displacement vectors, obtained from the Gaussian calculation. The polarizability calculations were performed in CFOUR. All calculations were done on the CCSD level, with the aug-cc-pVQZ-PP basis set. Energy calculations were performed in Gaussian at the same level of theory as was used for the vibrational analysis.

The vibrational contributions to the polarizabilities were calculated from the series of displaced geometry results by applying the Numerov--Cooley (NC) procedure \cite{Numerov1924, Cooley1961}, where the nuclear Schr\"odinger equation for the calculated potential curve is numerically solved, yielding the vibrational wavefunctions. The polarizabilities were then averaged over the vibrational wavefunction for the given vibrational mode and vibrational level. The differences between the NC-averaged values and the equilibrium values within the CCSD approach and the aug-cc-pVQZ-PP basis set were added as a vibrational correction to the electronic polarizabilities obtained at the higher level of theory.

\subsection{Two-step methodology validation}

As an independent check of our two-step approach to calculating the static and dynamic polarizabilities using two separate methods (DC-FF/ECP-LR), we carried out a comparison with the DC relativistic LR-CCSD method implemented in the quantum chemistry software package BAGH~\cite{dutta2023bagh}. In these calculations, a truncation of the virtual space based on frozen natural spinors (FNS) is used to reduce the computational cost. Chakraborty and co-workers recently implemented the perturbation-sensitive FNS method (FNS++)~\cite{chakraborty2025lowcost}, where a basis is constructed by diagonalizing perturbed MP2 densities obtained from perturbed first-order singles and doubles amplitudes. On this basis, all virtual spinors with occupation below 10$^{-5}$ are discarded.

\section{Results and Discussion}

\subsection{Electric dipole moment and static polarizability}\label{subsec:dipmom_statpol_results}
Table~\ref{tab:mu_alpha} contains the calculated molecular body-fixed dipole moment, and the parallel and perpendicular polarizabilities of BaOH, at the Hartree-Fock (HF), CCSD, and CCSD(T) levels of theory and using different basis sets. The calculations were performed within the X2C Hamiltonian, correlating 35 electrons and setting the virtual space cutoff at 20\,$E_\text{h}$.
Clearly, inclusion of electron correlation is crucial for the calculated dipole moment, as the HF values are 5 times lower than the CCSD/CCSD(T) results. For polarizabilities, the inclusion of correlation lowers the calculated values by 20--34$\%$. The opposite trends of the dipole moment and the dipole polarizability with the improving electron correlation can be explained by the increased ionic character of the Ba--OH bond upon introducing electron correlation.
An obvious consequence of the increase in bond ionicity is the increase in dipole moment, as seen in Table~\ref{tab:mu_alpha}. A significant decrease in polarizability of Ba$^+$ compared to neutral Ba is then reflected also in quasi-ionic Ba$^+$OH$^-$. Theoretical studies have found $\alpha(\text{Ba}) = 272(10)$~a.u.\cite{Schwerdtfeger2019}, $\alpha(\text{OH}) = 7.541$~a.u.\cite{Adamowicz1998}, $\alpha(\text{Ba}^+) = 124.15$~a.u.\cite{Iskrenova-Tchoukova2008}, and $\alpha(\text{OH}^-) = 47(4)$~a.u.\cite{Pluta1998}, indeed showing that $\alpha(\text{Ba}) + \alpha(\text{OH}) > \alpha(\text{Ba}^+) +\alpha(\text{OH}^-)$, consistent with the observed trends.

\begin{table}[!htbp]
    \centering
    \caption{Electric dipole moment ($\mu_z$), and parallel ($\alpha_\parallel$) and perpendicular ($\alpha_\perp$) static dipole polarizabilities (in a.u.) at the HF, CCSD, and CCSD(T) levels of theory, for various basis sets, using the X2C Hamiltonian and a CC energy active space of $\pm20.0$\,$E_\text{h}$.}
    \label{tab:mu_alpha}
    \begin{tabular}{lllll}
\hline
 & Basis set & HF  & CCSD  & CCSD(T) \\ \hline 
$\mu_z$ & s-aug-dyall.v2z & 0.179 & 0.564 & 0.588   \\
 & dyall.v3z       & 0.080 & 0.476 & 0.503     \\
 & s-aug-dyall.v3z & 0.106 & 0.528 & 0.565     \\
 & d-aug-dyall.v3z & 0.106 & 0.530 & 0.568     \\
 & t-aug-dyall.v3z & 0.106 & 0.530 & 0.569     \\
 & dyall.v4z       & 0.102 & 0.513 & 0.550     \\
 & s-aug-dyall.v4z & 0.103 & 0.524 & 0.566     \\ \hline
$\alpha_\parallel$ & s-aug-dyall.v2z & 245.0 & 206.8 & 206.1  \\
 & dyall.v3z        & 226.7 & 199.3 & 198.9  \\
 & s-aug-dyall.v3z  & 247.4 & 206.8 & 204.3  \\
 & d-aug-dyall.v3z  & 249.5 & 207.4 & 204.3  \\
 & t-aug-dyall.v3z  & 249.8 & 207.6 & 204.5  \\
 & dyall.v4z        & 243.9 & 204.6 & 203.7  \\
 & s-aug-dyall.v4z  & 249.8 & 207.2 & 202.9  \\ \hline
 $\alpha_\perp$ & s-aug-dyall.v2z & 455.7 & 320.9 & 311.7  \\
 & dyall.v3z        & 406.9 & 295.0 & 284.2  \\
 & s-aug-dyall.v3z  & 450.5 & 315.4 & 302.3  \\
 & d-aug-dyall.v3z  & 459.6 & 318.6 & 303.6  \\
 & t-aug-dyall.v3z  & 459.7 & 318.6 & 305.0  \\
 & dyall.v4z        & 451.1 & 315.5 & 298.1  \\
 & s-aug-dyall.v4z  & 459.0 & 318.5 & 300.6  \\ \hline
\end{tabular}

\end{table}

The calculated dipole moment is more sensitive to the basis set effects than the polarizability, but we found both properties to be converged to within 1\% at the singly-augmented v4z basis set level. We also found that $\alpha_{\perp}$ is significantly larger than $\alpha_{\parallel}$ because the electrons are less tightly bound in the direction perpendicular to the molecular axis and thus show a stronger response to the applied external electric field. This is also reflected in the fact that more augmentation functions were needed to converge the perpendicular polarizability values to the same 1\% level of precision as the parallel ones. The tests of augmentation were performed for the v3z quality basis set; calculations of perpendicular polarizability using doubly- and triply-augmented basis sets of quadruple cardinality were computationally intractable due to the reduced symmetry required for this property.

We have also investigated the relativistic and QED effects on the calculated dipole moment and polarizabilities, by comparing NR, X2C, DC, and (X2C-based) QED results (Table~\ref{tab:alpha_stat_hami}). These calculations were performed at the CCSD(T) level, using the s-aug-dyall.v3z basis set, correlating 35 electrons and setting the active space energy cut-offs at $\pm20.0$\,$E_\text{h}$.
As shown in Table~\ref{tab:alpha_stat_hami}, the relativistic values of the polarizabilities are significantly lower than the NR ones. This is likely due to the relativistic contraction of the s-type orbitals of the barium atom, which reduces the size of the valence orbitals, thus reducing the polarizability. Additionally, we see that the results obtained within the X2C Hamiltonian are in excellent agreement with DC values, justifying the use of the X2C approximation for these properties.

\begin{table}[!hbpt]
    \centering
    \caption{Electric dipole moment and parallel ($\alpha_\parallel$) and perpendicular ($\alpha_\perp$) static polarizabilities of BaOH in a.u., calculated within NR, X2C, DC, and (X2C-based) QED frameworks, at CCSD(T) level of theory, with $\pm20.0$\,$E_\text{h}$ active space and s-aug-dyall.v3z basis set.}
    \label{tab:alpha_stat_hami}
    \begin{tabular}{lllll}
    \hline
                    & NR       & X2C    & DC    & QED   \\ \hline 
 $\mu_z$            & 0.391    & 0.565  & 0.566 & 0.562 \\
 $\alpha_\parallel$ & 239.5    & 204.3  & 204.1 & 204.8 \\
 $\alpha_\perp$     & 375.3    & 302.3  & 301.9 & 302.0 \\
 \hline
\end{tabular}
\end{table}

In Table~\ref{tab:alpha_stat_AS}, the dependence of the dipole moment and the polarizabilities on the size of the CC active space is shown, based on X2C-CCSD(T) calculations with dyall.ae3z basis set. The dependence of the results on the active space is minimal, as expected for valence properties. We thus conclude that using the smaller active space is sufficiently accurate.

\begin{table}[!hbpt]
    \centering
    \caption{Electric dipole moment and parallel ($\alpha_\parallel$) and perpendicular ($\alpha_\perp$) static polarizabilities (a.u.), calculated at the CCSD(T) level with active spaces between --20\,$E_\text{h}$ and 20\,$E_\text{h}$ (35 correlated electrons), --200\,$E_\text{h}$ and 200\,$E_\text{h}$ (59 correlated electrons), and --2000\,$E_\text{h}$ and 2000\,$E_\text{h}$ (65 correlated electrons), using the dyall.ae3z basis set.}
    \label{tab:alpha_stat_AS}
    \begin{tabular}{cccc}
 \hline
                           & $\pm$20\,$E_\text{h}$ & $\pm$200\,$E_\text{h}$ & $\pm$2000\,$E_\text{h}$ \\ 
 \hline
$\mu_z$            & 0.494      & 0.496       & 0.496              \\
$\alpha_\parallel$ & 197.8      & 197.4       & 197.4              \\
$\alpha_\perp$     & 280.1      & 279.0       & 279.2              \\
\hline
\end{tabular}
\end{table}

The final results were obtained by increasing the level of theory, the basis set cardinality and augmentation, and CC active space until convergence to within about 1\% was observed; these investigations were done independently for each computational parameter.
The final recommended values of the body-fixed electric dipole moment and the static polarizabilities $\mu_z$, $\alpha_{\parallel,\text{DC}}(0)$, and $\alpha_{\perp,\text{DC}}(0)$, respectively, were obtained by taking the results based on the s-aug-v4z basis set, X2C Hamiltonian and $\pm 20.0$\,$E_\text{h}$ active space (base values) and correcting these for the differences between calculations with: (i) DC and X2C Hamiltonians; (ii) X2C results with and without inclusion of effective QED potentials; (iii) active space cutoff energies of $\pm20$\,$E_\text{h}$ and $\pm2000$\,$E_\text{h}$; and (iv) s-aug-dyall.v3z and t-aug-dyall.v3z basis sets. Two additional corrections were added, so that: (v) the static polarizabilities were shifted by --1.3\%, corresponding to the difference between the static polarizability of the barium atom, determined in Ref.~\citenum{Chattopadhyay2014} with the Dirac--Coulomb--Breit (DCB) and DC Hamiltonians; and (vi) 
the vibrational effects were accounted for as the sum of the shifts in the property's value, compared to the equilibrium value, for the ground state of each of the four vibrational modes. For the bending modes, the perpendicular polarizability shift was taken as the sum of the two (now different) polarizability shifts along the $x$ and $y$ axes, perpendicular to the molecular axis (see Table~\ref{tab:alpha_stat_mu_contr}).

 Two-electron Coulomb integrals of ($SS$$\mid$$SS$) type were omitted from the calculations and replaced by simple Coulombic energy corrections\cite{Visscher1997-SSSS}, which is the standard procedure in the DIRAC code, since we have found that including them has a negligible effect on the dipole moment and polarizabilities.

\begin{table}[!hbpt]
    \centering
    \begin{threeparttable}
    \caption{Base values (s-aug-dyall.v4z basis set, X2C Hamiltonian, --20\,$E_\text{h}$ to 20\,$E_\text{h}$ energy active space, equilibrium geometry) and corrections to the calculated dipole moment and static polarizability,  alongside the scheme by which the latter were determined. The final values were obtained by adding the base values and the corrections.}
    \label{tab:alpha_stat_mu_contr}
    \setlength{\tabcolsep}{3pt}
    \begin{tabular}{lrrrl}
    \hline
  Contribution  & $\mu_z$ & $\alpha_{\parallel,\text{DC}}(0)$ & $\alpha_{\perp,\text{DC}}(0)$ & Scheme          \\ \hline 
  Base value    & 0.566   & 202.86 & 300.62 &                   \\
  Hamiltonian   & 0.001   & --0.21 & --0.49 & DC -- X2C         \\
  QED           & --0.003 & 0.49   & --0.36 & QED -- X2C        \\
  Active space  & 0.002   & --0.36 & --0.93 & $\pm2000 - \pm20$ \\
  BSA\tnote{a}  & 0.004   & 0.21   & 2.66   & t-aug -- s-aug    \\
  Breit         & --      & --2.63 & --3.90 & Ba: DCB -- DC     \\
  VC\tnote{b} & 0.014 & 0.41   & --0.19 & (000) -- Equil.       \\ \hline
  Final value   & 0.583   & 200.76 & 297.40 & $\Sigma_i$        \\
 \hline
\end{tabular}
\begin{tablenotes}
    \item[a] Basis set augmentation (BSA).
    \item[b] Vibrational correction (VC).
\end{tablenotes}
\end{threeparttable}
\end{table}

\subsubsection{Uncertainty analysis}

We have carried out an evaluation of the uncertainty of the predicted dipole moment and polarizabilities, based on the estimates of the effect of incompleteness of the treatment of various computational parameters. 
The uncertainty corresponding to each computational parameter was analyzed separately and was taken as the difference between the results obtained for calculations performed on the highest and second-highest level of treatment of the corresponding parameter. The total uncertainty $\sigma_{\text{stat}}$ was determined by taking the square root of the sum of the squares (SRSS) of the individual uncertainties, assuming mutual independence:

\begin{itemize}
    \item uncertainty due to using an incomplete basis set, estimated as half the difference between the s-aug-dyall.v3z and the s-aug-dyall.v4z basis set values,
    \item basis augmentation uncertainty is estimated as the difference between the t-aug-dyall.v3z and d-aug-dyall.v3z basis sets,
    \item uncertainty associated with omitting higher-order excitations beyond perturbative triples in the CC procedure is estimated conservatively as 20\% of the triples contribution, i.e., relative difference between the static polarizabilities calculated at the CCSD and CCSD(T) levels of theory
    (in previous studies, higher excitation contribution to polarizability was found to be around 10\% of the triples contribution~\cite{Jerabek2018,Sahoo2018}),
    \item active space incompleteness uncertainty is taken as the difference between values obtained using an active space cutoff of $\pm 200$\,$E_\text{h}$ and of $\pm 2000$\,$E_\text{h}$,
    \item uncertainty due to the atomic estimate of the Breit correction is estimated as 1\% of the total polarizability, based on earlier research on the polarizability of Ba and some other atoms where Breit contributions to polarizabilities of 0.5\%--1.5\% were found \cite{Chattopadhyay2014, Cheng2025, Safronova2012, Kumar2020}.
\end{itemize}

The partial relative and absolute uncertainties are shown in Table~\ref{tab:pol_sigma}, alongside the scheme whereby they were determined. The largest contribution to the total uncertainty is the omission of higher-order CC excitations for the dipole moment and perpendicular polarizability, and the Breit estimate for the parallel polarizability.

\begin{table*}[!htbp]
    \centering
    \caption{Partial and total uncertainties of the electric dipole moment and static polarizabilities.}
    \label{tab:pol_sigma}
    \begin{tabular}{@{\extracolsep{4pt}}llllllll@{}}
    \hline
& \multicolumn{2}{c}{$\mu_{z}$} & \multicolumn{2}{c}{$\alpha_\parallel$} & \multicolumn{2}{c}{$\alpha_\perp$} & \\ \cline{2-3} \cline{4-5} \cline{6-7}
Source & $\sigma_{i}$ (a.u.) & $\sigma_{i}$ (\%) & $\sigma_{i}$ (a.u.) & $\sigma_{i}$ (\%) & $\sigma_{i}$ (a.u.)& $\sigma_{i}$ (\%) & Scheme \\ \hline 
Basis cardinality  & 0.0003 & 0.06 & 0.73 & 0.36 & 0.85 & 0.29 & 0.5(s-v4z $-$ s-v3z)   \\
Basis augmentation & 0.0004 & 0.07 & 0.21 & 0.10 & 1.40 & 0.47 & t-v3z $-$ d-v3z        \\
Active space       & 0.0005 & 0.09 & 0.05 & 0.03 & 0.24 & 0.08 & $\pm2000\,E_\text{h}-\pm200\,E_\text{h}$  \\
Higher excitations & 0.0085 & 1.45 & 0.86 & 0.42 & 3.55 & 1.19 & 0.2(CCSD(T) $-$ CCSD)  \\
QED                & 0.0030 & 0.52 & 0.48 & 0.24 & 0.36 & 0.12 & QED -- DC              \\
Breit              & 0.0058 & 1.00 & 2.01 & 1.00 & 2.97 & 1.00 & Ba atom                \\ \hline
Total              & 0.0108 & 1.84 & 2.36 & 2.09 & 4.93 & 1.66 & SRSS($\sigma_i$)       \\ \hline
\end{tabular}
\end{table*}

\subsubsection{Final values}

The final vibrationally corrected body-fixed electric dipole moment and the polarizabilities, including the uncertainties, are given in Table~\ref{tab:final_results}. The agreement of the calculated dipole moment $\mu_z=0.583(11)$ a.u. with the measured value $\mu_z=0.563(16)$ a.u. \cite{Frey2011} supports our theoretical predictions for the static parallel and perpendicular polarizabilities, where no experimental results are available yet, are reliable.

\subsection{Dynamic polarizability}\label{subsec:dynpol_results}

Dynamic polarizabilities were calculated using the ECP approximation in CFOUR (Section~\ref{sec:comp-dyn}). In order to assess the reliability of the ECP results, in Table~\ref{tab:alpha_stat_shift} we compared the static polarizabilities calculated with DC and ECP, on the CCSD level. The difference due to the use of the ECP approximation is minimal, and might also be partially due to the different basis set used in the two calculations. 
Nevertheless, as we know that the level of theory for the $\alpha_{\text{DC}}(0)$ calculations in DIRAC is significantly higher (large uncontracted basis sets, inclusion of perturbative triple excitations and spin-orbit effects by means of the X2C relativistic Hamiltonian), we therefore shifted the data sets of dynamic polarizability values such that the static limit $\alpha_{\text{ECP}}(0)$ coincides with the final result for the static polarizability $\alpha_{\text{DC}}(0)$.

\begin{table}[!hbpt]
    \centering
    \caption{Parallel ($\parallel$) and perpendicular ($\perp$) static polarizabilities in a.u. The DC values were obtained with the s-aug-dyall.v3z basis set, and the ECP values are the static limits obtained with the aug-cc-pVTZ-PP basis set.}
    \label{tab:alpha_stat_shift}
    \begin{tabular}{cll}
    \hline
                    & $\alpha_\text{DC}(0)$    & $\alpha_\text{ECP}(0)$ \\ \hline 
$\parallel\;\;\;\;$ & 206.5                    & 210.5                  \\
$\perp\;\;\;\;$     & 314.9                    & 316.9                  \\ \hline
\end{tabular}
\end{table}

For each set of calculated dynamic polarizability values $\alpha(\omega)$, a parametrized fit of the form $\alpha(\omega)=\alpha(0)+\frac{b\omega}{\omega-c}$ was used i) for interpolation to an arbitrary frequency of the electric field $\omega$, and ii) for robust uncertainty estimation (see below).
In this model, $c$ represents an asymptote corresponding to the transition frequency for an excited state. Due to the symmetry of the molecule and the electric field, the asymptotes for the parallel and perpendicular components of the polarizability correspond to the $B^2\Sigma \leftarrow X^2\Sigma$ and $A^2\Pi \leftarrow X^2\Sigma$ transitions, respectively. The $A'^2\Delta \leftarrow X^2\Sigma$ transition is a quadrupole transition and does not manifest as an asymptote in the dipole polarizability curves. The effect of triple CC excitations is included indirectly by correcting the fitting parameter $c$ by $\Delta$(T) correction calculated as the difference between separately calculated transition energies based on single-point energy calculations at CCSD and CCSD(T) levels of theory.
The calculated $A^2\Pi \leftarrow X^2\Sigma$ transition energies at CCSD and CCSD(T) levels were found to be 11520~cm$^{-1}$ and 11641~cm$^{-1}$ respectively, comparing well to the experimental value of 11763 cm$^{-1}$ \cite{Tandy2009}.

The effect of vibration on the dynamic polarizability was determined in the same manner as for the static polarizability.

\subsubsection{Uncertainty analysis}
The parameter $\alpha(0)$ of the fit equation is set to $\alpha_{\text{DC}}(0)$ calculated in Section~\ref{subsec:dipmom_statpol_results}.
For the two remaining free fitting parameters $b$ and $c$, uncertainties were assigned by combining uncertainties due to the different computational parameters, similar to the procedure employed in Section~\ref{subsec:dipmom_statpol_results}:

\begin{itemize}
    \item basis set cardinality uncertainty is taken as half of the difference between 5Z and QZ results,
    \item basis set augmentation uncertainty is set to the difference between the doubly- and singly-augmented basis sets results,
    \item uncertainty associated with the non-linear least-squares fit,
    \item uncertainty associated with omitting higher-order excitations beyond perturbative triples in the CC procedure is estimated as 20\% of the triples (based on the difference between the $A^2\Pi \leftarrow X^2\Sigma$ transition energies calculated at CCSD and CCSD(T) levels of theory). The same uncertainty was assigned to both $\alpha_\parallel$ and $\alpha_\perp$ since calculating the $B^2\Sigma$ state is not possible within the same single-reference computational scheme.
\end{itemize}

The individual uncertainty sources were again combined by means of SRSS into the total uncertainties of parameters $b$ and $c$. These were then used to determine the uncertainty in the dynamic polarizability $\sigma_{\text{dyn}}$ at any desired frequency through standard uncertainty propagation rules.

The values for fitting parameters $b$ and $c$ for different basis sets are listed in Table~\ref{tab:dyn_pol_fit_param}, together with the experimental values for the relevant electronic transitions. The fitted transition energies $c$ align well with the experiment \cite{Tandy2009}.

\begin{table*}[!htbp]
    \centering
    \begin{threeparttable}
    \caption{Values for fitting parameters $b$ and $c$ (in a.u.) for the parallel and perpendicular dynamic polarizability curves for various basis sets. $c$ is also given in cm$^{-1}$ for comparison with experimental transition frequencies. The highlighted values are used in the determination of the final recommended values.}
    \label{tab:dyn_pol_fit_param}
    \begin{tabular}{lllll}
    \hline
   &  & $b$  & $c$ & $c\;(\text{cm}^{-1})$ \\ \hline
$\alpha_\parallel\;\;\;\;$ & aug-cc-pVQZ-PP    & 64.8  & 0.059586 & 13077    \\
                           & 2-aug-cc-pVQZ-PP  & 54.4  & 0.058170 & 12765    \\
                           & aug-cc-pV5Z-PP    & \textbf{63.2}  & \textbf{0.059523} & 13064    \\ 
                           & Exp. ($B^2\Sigma \leftarrow X^2\Sigma$ )  &  &  & 13200\tnote{a} \\ \hline 
$\alpha_\perp\;\;\;\;$     & aug-cc-pVQZ-PP    & 111.5 & 0.053092 & 11652    \\
                           & 2-aug-cc-pVQZ-PP  & 107.9 & 0.053421 & 11724    \\
                           & aug-cc-pV5Z-PP    & \textbf{106.4} & \textbf{0.053229} & 11682    \\
                           & Exp. ($ A^2\Pi\leftarrow X^2\Sigma$)  &  &  & 11763\tnote{bc}    \\
  \hline
    \end{tabular}
    \begin{tablenotes}
    \item[a] Ref. \cite{KinseyNielsen_1986}.
    \item[b] Ref. \cite{Tandy2009}.
    \item[c] Average of $A^2\Pi_{1/2}\leftarrow X^2\Sigma$ and $A^2\Pi_{3/2}\leftarrow X^2\Sigma$.
    \end{tablenotes}
    \end{threeparttable}
\end{table*}

In Table~\ref{tab:dyn_pol_sigmas}, the uncertainties in the fitting parameters due to the incompleteness of the computational parameters are shown, alongside the scheme whereby they were determined. The uncertainties of the dynamic polarizabilities are shown as well, determined by propagating the corresponding $\sigma_b$ and $\sigma_c$ through the \mbox{aug-cc-pV5Z-PP} basis set fit.

\begin{table*}[!htbp]
    \centering
    \caption{Uncertainties of the fitting parameters $b$ and $c$, alongside the scheme whereby they were determined, and of the dynamic polarizabilities $\alpha$ at $\lambda=1064$ nm ($\omega=282$ THz), which were determined by propagating the corresponding $\sigma_b$ and $\sigma_c$ through the aug-cc-pV5Z-PP data set fit.}
    \label{tab:dyn_pol_sigmas}
    \setlength{\tabcolsep}{4pt}
    \begin{tabular}{@{\extracolsep{3pt}}llllllllll@{}}
    \hline
& \multicolumn{4}{c}{$\alpha_\parallel$} & \multicolumn{4}{c}{$\alpha_\perp$} & \\ \cline{2-5} \cline{6-9}
Source      & $\sigma_b$      & $\sigma_c$ & $\sigma_\alpha$ (a.u.) & $\sigma_\alpha$ (\%) & $\sigma_b$ & $\sigma_c$ &  $\sigma_\alpha$ (a.u.) & $\sigma_\alpha$ (\%) & Scheme \\ \hline
Basis cardinality    &  \phantom{0}0.8 & 0.00003 & \phantom{0}2.2 & 0.60 & 2.59           & 0.00007 &             11 & 1.49 & 0.5(aug-5z $-$ aug-qz) \\
Basis augmentation    &            10.4 & 0.00142 &             30 & 8.22 & 3.66           & 0.00033 &             21 & 2.84 & 2-aug-qz $-$ aug-qz    \\
Higher excitations & \phantom{00} -- & 0.00006 & \phantom{0}1.1 & 0.30 & \phantom{00}-- & 0.00006 & \phantom{0}3.2 & 0.43 & 0.2(CCSD(T) $-$ CCSD)  \\ 
Fitting     &  \phantom{0}1.8 & 0.00008 & \phantom{0}5.0 & 1.38 & 2.64           & 0.00007 &             11 & 1.49 & aug-5z fit $\sigma$     \\ \hline
Total       &            10.5 & 0.00142 &           30.5 & 8.36 & 5.20           & 0.00035 &             26 & 3.56 & SRSS                   \\ \hline
\end{tabular}
\end{table*}

\subsubsection{Final values}

\begin{figure}[!hbpt]
    \centering
    \includegraphics[width=\linewidth]{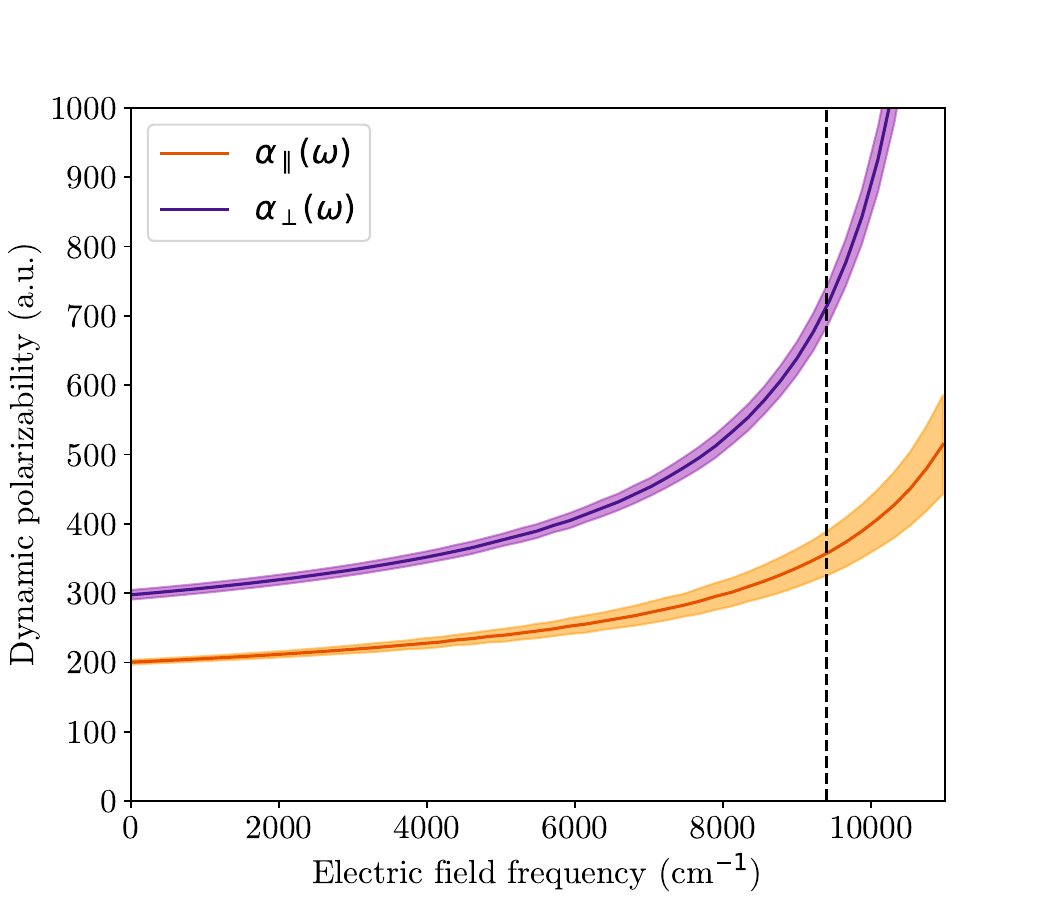}
    \caption{Parallel and perpendicular components of the dynamic polarizability (solid lines) with uncertainty bounds (shaded regions) as a function of the frequency of the external electric field. The dashed line corresponds to the 1064 nm laser.}
    \label{fig:dyn_pol}
\end{figure}

Figure~\ref{fig:dyn_pol} shows the final parallel and perpendicular dynamic polarizabilities with confidence bands as functions of the laser frequency. The dynamic polarizabilities increase asymptotically as the laser frequency approaches the electronic transition frequency. Note that the states do not exhibit spin-orbit splitting here since all dynamic polarizability calculations only take scalar relativistic effects into account.

The final results for the parallel and perpendicular components of the polarizability tensor of BaOH at equilibrium geometry, and in the (000) and (010) vibrational states at the static limit and at the frequency of the laser are summarized in Table~\ref{tab:final_results}. The molecular vibration has minimal effect on the polarizability.

\begin{table}[H]
    \centering
    \caption{Final results for the body-fixed electric dipole moment and parallel ($\alpha_\parallel$) and perpendicular ($\alpha_\perp$) static and dynamic polarizabilities of BaOH (in a.u.), in the equilibrium (eq.) electronic, and in the (000) and (010) vibrational states.}
    \label{tab:final_results}
    \begin{tabular}{llll}
    \hline
  & State & $\omega=0$ & $\omega=282$ THz \\ \hline
$\mu_z$            & (000)  & 0.583(11) & --      \\
$\alpha_\parallel$ & eq.    & 200.3(24) & 357(31) \\
                   & (000)  & 200.8(24) & 358(31) \\
                   & (010)  & 201.1(24) & 360(31) \\
$\alpha_\perp$     & eq.    & 298(5)    & 714(29) \\
                   & (000)  & 297(5)    & 715(29) \\
                   & (010)  & 297(5)    & 718(29) \\ \hline
\end{tabular}
\end{table}

\subsubsection{DC-LRCCSD comparison}

Table~\ref{tab:BAGH_results} compares the static and dynamic polarizabilities obtained using the composite scheme presented above with the one obtained when the unrelaxed Dirac--Coulomb four-component linear response coupled cluster with single and double excitations (DC-LRCCSD) method is applied using the program BAGH \cite{chakraborty2025lowcost}.
For the static polarizability, the comparison is made at the vibrationally uncorrected CCSD level using s-aug-dyall.v4z basis sets with $\pm$20\,$E_\text{h}$ correlation space (correlating 35 electrons). Excellent agreement is found between the two methods. The residual differences of 2--4\% can be attributed mostly to the orbital relaxation effects (present in FF, neglected in LR) and partially to the differences in the treatment of the virtual correlation space. In case of BAGH calculations, the FNS++ approach reduced the virtual space by 64\% with respect to the canonical space with the 20\,$E_\text{h}$ cutoff. The error introduced by this approximation is expected to amount to only a few tenths of a.u.~\cite{chakraborty2025lowcost}, i.e., at the level $\sim0.1\%$ on the relative scale.

In case of the LR-ECP-CCSD dynamical polarizability calculations in CFOUR, the full virtual space and the aug-cc-pV5Z-PP basis set were used. The latter is a contracted basis, and it is thus closer in terms of quality to the fully uncontracted s-aug-dyall.v4z basis set used in the other calculations than the aug-cc-pVQZ-PP basis set would be.

\begin{table}[H]
    \centering
    \caption{Comparison of the static and dynamic ($\omega$~=~282~THz) polarizabilities (a.u.) of BaOH calculated at CCSD/aQZ level between the two-step DC/ECP and one-step DC-FNS++ approaches described in Section~\ref{comp}.}
    \label{tab:BAGH_results}
    \begin{tabular}{@{\extracolsep{4pt}}llll}
    \hline
            &                  &  DC/ECP & DC-FNS++ \\ \hline
$\parallel$ & $\alpha(0)$      &  206.9  & 202.0    \\
            & $\alpha(\omega)$ &  368.9  & 352.3    \\
$\perp$     & $\alpha(0)$      &  318.1  & 308.1    \\
            & $\alpha(\omega)$ &  756.1  & 743.4    \\ \hline
    \end{tabular}
\end{table}

\subsection{Experimental implications}

The lowest dipole-allowed transition in BaOH for light that is polarized perpendicular (parallel) to the molecular axis lies at $871\,\text{nm}$ ($\sim750\,\text{nm}$). As a result, the value of $\alpha_\perp$ ($\alpha_\parallel$) at $\lambda=1064\,\text{nm}$ (see Table~\ref{tab:final_results}) deviates from the static case. The calculated polarizabilities given in Table~\ref{tab:final_results} compare well with \textit{ab initio} calculations on CaOH \cite{Hallas2023} (reportedly giving $\alpha_\perp=234.6\;\text{a.u.}$~and $\alpha_\parallel=142.6\;\text{a.u.}$~at $\lambda=1064\,\text{nm}$), indicating the feasibility of using similar trapping configurations.

In the case of BaOH, the results in Table~\ref{tab:final_results} indicate the possibility of creating a $\sim1\,\text{mK}$ deep ODT by using $\sim7\,\text{W}$ of laser power at a trap waist of $\sim25\,\mu\text{m}$. Alternatively, a deeper and/or wider trap may also be formed using the enhanced field inside an optical cavity. For a deep ODT ($\gtrsim10\,\text{mK}$), residual photon scatterings due to the off-resonant trapping laser may become significant. Following the formalism by Grimm \textit{et al.} \cite{Grimm1999}, the scattering rate can be estimated through
\begin{align}
    \gamma_{\text{scat}}\approx\frac{U}\hbar\frac\Gamma\Delta,
\end{align}
where $U$ represents the trap depth, $\hbar$ the reduced Planck constant, $\Gamma$ the transition linewidth, and $\Delta$ the detuning of the trapping light from that transition. Assuming a linewidth of $\Gamma=3\,\text{MHz}$ for the lowest dipole-allowed transition in BaOH at $\lambda=871\,\text{nm}$ ($A^2\Pi_{1/2}\leftarrow X^2\Sigma$), this results in a $1\,\text{mK}$ deep trap scattering a photon every second, or $10$ times per second for a $10\,\text{mK}$ deep trap. To suppress this effect, one can either reduce the trap depth, which requires more extensive cooling of the BaOH molecules, or increase the detuning.

\section{Conclusion}
The NL-eEDM collaboration recently proposed an eEDM experiment making use of trapped BaOH molecules in the (010) vibrational state. Both the static polarizability and the dynamic polarizability at $\lambda=1064\,\text{nm}$ (commonly used for trapping) were calculated using relativistic coupled-cluster theory. Uncertainties were assigned to these results by systematically varying the relevant computational parameters.
The final static polarizabilities were derived from the s-aug-dyall.v4z basis sets results, and were vibrationally corrected. For the dynamic polarizability, we devised a procedure to constrain the uncertainties by making use of the more accurately determined static limit. The calculated polarizabilities can be used for designing and optimizing next-generation precision experiments with trapped polyatomic molecules. \newline

\begin{acknowledgments}
We thank the Center for Information Technology of the University of Groningen for their support and for providing access to the H\'abr\'ok high-performance computing cluster.

The work of LFP, BS, and SH was supported by project number VI.C.212.016 financed by the Dutch Research Council (NWO). The work of SH, AB, and IAA was supported by the ENW-M2 NWO grant OCENW.M.21.098. The work of SH, AB, and EP is supported by ENW-XL NWO grant OCENW.XL21.XL21.074. IAA acknowledges partial support from FONCYT through grants PICT-2021-I-A-0933 and PICT-2020-SerieA-0052, and from CONICET through grant PIBAA-2022-0125CO. LFP acknowledges partial support from the Scientific Grant Agency of the Slovak Republic (project 1/0254/24).

IAA would like to express his special gratitude to the VSI Institute for Particle Physics and Gravity at the University of Groningen for its warm hospitality during his recent stay, where stimulating discussions for this study took place.
\end{acknowledgments}

\bibliography{bibliography}

\end{document}